\documentclass[aps,prx,reprint,superscriptaddress,showpacs,showkeys]{revtex4-2}
\usepackage{graphicx}
\usepackage{siunitx}

\begin{document}

\title{Suppression of dark-state polariton collapses in a cold-atom quantum memory}

\author{Katja Gosar}
\email[]{katja.gosar@ijs.si}
\affiliation{Jo\v{z}ef Stefan Institute, Jamova 39, SI-1000 Ljubljana, Slovenia}
\affiliation{Faculty of Mathematics and Physics, University of Ljubljana, Jadranska 19, SI-1000 Ljubljana, Slovenia}

\author{Vesna Pirc Jev\v{s}enak}
\affiliation{Jo\v{z}ef Stefan Institute, Jamova 39, SI-1000 Ljubljana, Slovenia}
\affiliation{Faculty of Mathematics and Physics, University of Ljubljana, Jadranska 19, SI-1000 Ljubljana, Slovenia}

\author{Tadej Me\v{z}nar\v{s}i\v{c}}
\affiliation{Jo\v{z}ef Stefan Institute, Jamova 39, SI-1000 Ljubljana, Slovenia}
\affiliation{Faculty of Mathematics and Physics, University of Ljubljana, Jadranska 19, SI-1000 Ljubljana, Slovenia}

\author{Samo Begu\v{s}}
\affiliation{Faculty of Electrical Engineering, University of Ljubljana, Tr\v{z}a\v{s}ka cesta 25, SI-1000 Ljubljana, Slovenia}

\author{Tomasz Krehlik}
\affiliation{Photonics Department, Jagiellonian University, \L{}ojasiewicza 11, 30-348 Krak\'{o}w, Poland}

\author{Du\v{s}an Ponikvar}
\affiliation{Jo\v{z}ef Stefan Institute, Jamova 39, SI-1000 Ljubljana, Slovenia}
\affiliation{Faculty of Mathematics and Physics, University of Ljubljana, Jadranska 19, SI-1000 Ljubljana, Slovenia}

\author{Erik Zupani\v{c}}
\affiliation{Jo\v{z}ef Stefan Institute, Jamova 39, SI-1000 Ljubljana, Slovenia}
\affiliation{Faculty of Natural Sciences and Engineering, University of Ljubljana, A\v{s}ker\v{c}eva cesta 12, SI-1000 Ljubljana, Slovenia}

\author{Peter Jegli\v{c}}
\email[]{peter.jeglic@ijs.si}
\affiliation{Jo\v{z}ef Stefan Institute, Jamova 39, SI-1000 Ljubljana, Slovenia}
\affiliation{Faculty of Mathematics and Physics, University of Ljubljana, Jadranska 19, SI-1000 Ljubljana, Slovenia}

\date{\today}

\begin{abstract}
We observe dark-state polariton collapses and revivals in a quantum memory based on electromagnetically induced transparency on a cloud of cold cesium atoms in a magnetic field. Using $\sigma^+$ polarized signal and control beams in the direction of the magnetic field, we suppress the dark-state polariton collapses by polarizing the atoms towards one of the stretched Zeeman states and optimizing the frequency detuning of the control beam. In this way, we demonstrate a quantum memory with only partial dark-state polariton collapses, making the memory usable at any storage time, not only at discretized times of revivals. We obtain storage time of more than \SI{400}{\micro s}, which is ten times longer than what we can achieve by trying to annul the magnetic field.
\end{abstract}

\pacs{}

\maketitle

\section{Introduction}
The ability to coherently store light and to recall it at a later time is essential for quantum communication \cite{Duan2001,Bhaskar2020}. Such quantum memories have been subject to a lot of research in recent years \cite{Heshami2016,Lvovsky2009}, and the use of electromagnetically induced transparency (EIT) on hot and cold atoms has proven to be a very promising method \cite{Ma2017,Fleischhauer2005}. EIT occurs when two laser beams that form a $\Lambda$-type system are shone onto a dense cloud of atoms. These beams drive a two-photon transition from the ground state to the storage state via an excited state. The signal beam couples the ground state to an excited state, while the control beam couples the storage state and excited state. The signal beam is much weaker than the control beam. Under these conditions, the absorption of the signal beam is greatly reduced, and the refractive index undergoes a steep variation at the resonance frequency. This leads to a strong reduction of the group velocity of the signal beam, causing a phenomenon called slow light \cite{Bayford2023,Lukin1997,Hau1999}. 

A pulse of the signal beam, slowed down by EIT, can be described as a quasi-particle called a dark-state polariton (DSP) \cite{Fleischhauer2000, finkelstein2023practical, derose2023producing}. It has an electromagnetic component and an atomic component. While the signal pulse is slowly propagating through the atoms, the stronger beam, called the control beam, can be adiabatically turned off. This causes the electromagnetic part of the DSP, along with the group velocity of the signal pulse, to be reduced to zero. The information of the signal pulse is thus stored in the spin coherence between the ground state and the storage state. This coherence is called a spin-wave and evolves temporally with a frequency $\omega_{sw} = \omega_s - \omega_g$, where $\hbar\omega_s$ and $\hbar\omega_g$ are the energies of the storage and ground states. After a desired time, adiabatically turning the control beam back on transfers the information from the atomic component of the DSP back into the electromagnetic component and the signal pulse is restored \cite{jenkins2006theory,Hsu2013,peters2009optimizing}. 

For the ideal quantum memory, a high efficiency and a long time of storage are desired. The latter is limited by dephasing of the atomic coherence due to the atomic motion \cite{Carvalho2004,Mewes2005} and any magnetic field gradients that might be present \cite{Happer1972}. While annulling stray fields is hard, especially in cold atom experiments, deliberately turning on a strong perpendicular homogeneous magnetic field has been shown to actually improve the lifetime of cold atom-based quantum memories \cite{Moretti2010}. Moreover, even a weak residual magnetic field may cause the atomic states to undergo Zeeman splitting and the $\Lambda$-systems are no longer degenerate. Consequently, once we turn off the control beam to store the signal pulse, many spin-waves with different energies $\hbar\omega_{sw}$ are formed. Because these spin-waves evolve with different frequencies, they interfere with each other. Depending on when we turn the control beam back on, this causes collapses and revivals of the amplitude of the retrieved light pulse as a function of storage time. If the magnetic field is weak, as residual fields in cold atom experiments usually are, the time between revivals is large  \cite{matsukevich2006observation,jenkins2006theory}. This may result in only the initial collapse being visible, as the consequent revivals are further than the intrinsic lifetime of the memory allows. Therefore, the effective lifetime of the quantum memory is much shorter than if there was no magnetic field present. If, however, we turn on a stronger magnetic field, the time between revivals decreases and much longer lifetimes are achievable, limited now mostly by just the atomic motion.

There remains one major challenge. Due to these collapses, the quantum memory is, in a way, discretized. The question, therefore, is how to reduce these collapses so that the quantum memory can be used for all storage times.

In this article, we first show how the effective lifetime of the quantum memory can be improved by applying a homogeneous magnetic field on unpolarized cesium cold atoms. Then we show how polarizing the atoms in an even stronger magnetic field suppresses the storage collapses. Lastly, we demonstrate the use of frequency selectivity to decrease the collapses even further and overall increase the quantum memory lifetime tenfold.

\begin{figure}
\includegraphics[width=\linewidth]{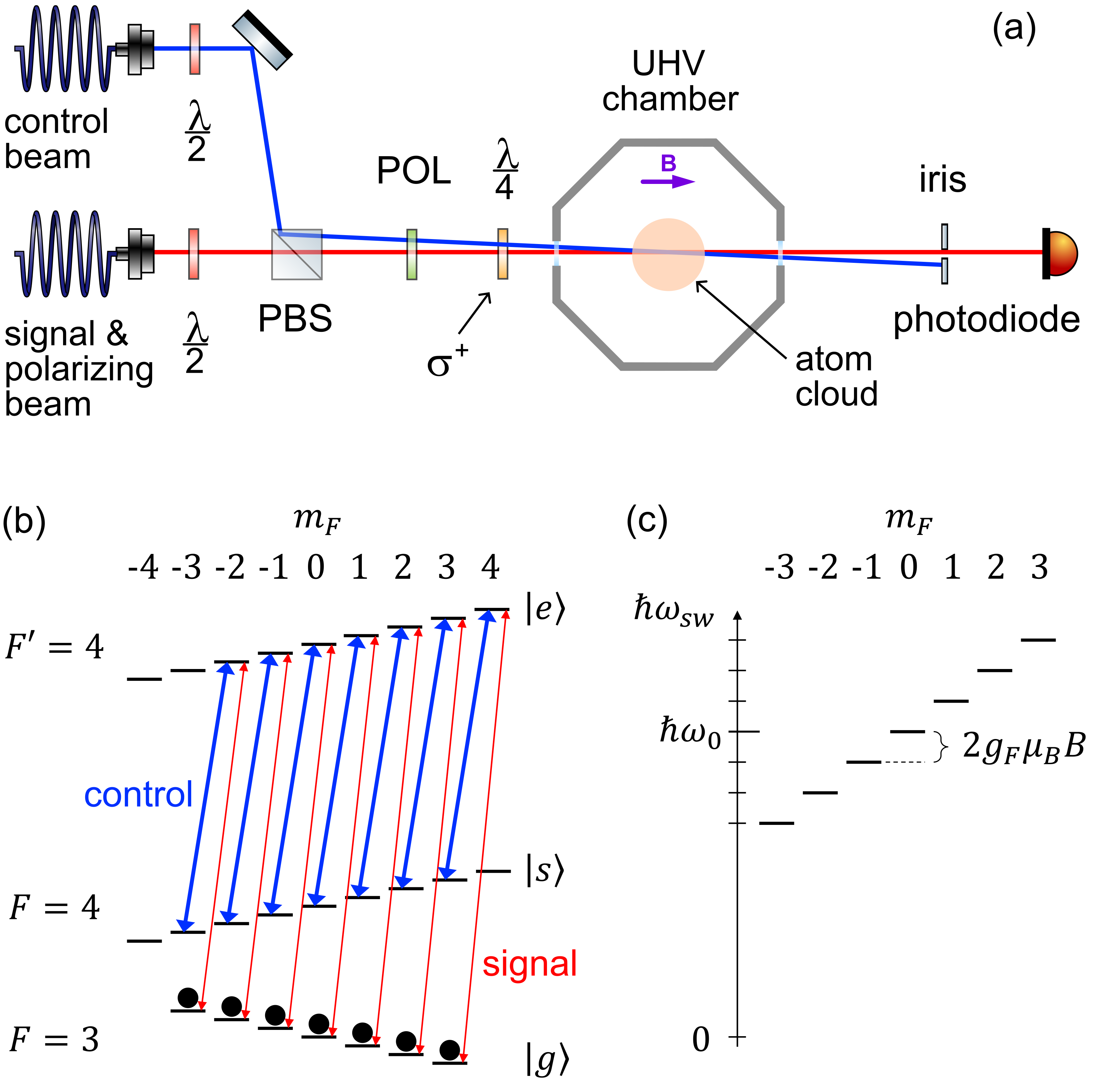}
\caption{
(a) Experimental setup for Section \ref{polarized}. 
We combine the control and signal beam on a polarizing beam splitter (PBS) and set both polarizations to $\sigma^+$ with a polarizer (POL) and a quarter-wave plate. The two beams travel at an angle of $\sim0.5^\circ$ and intersect at the position of the atomic cloud in the ultra-high vacuum chamber. On the other side of the chamber, we block the control light with an iris and measure the intensity of the signal beam with a photodiode.
(b) Energy levels of Cs D2 transition used for EIT. The control beam drives the transition $\vert F = 4\rangle \rightarrow \vert F' = 4\rangle$ and the signal beam is on the $\vert F = 3 \rangle \rightarrow \vert F' = 4 \rangle$ transition. In the presence of a magnetic field, the $m_F$ levels are not degenerate because of Zeeman splitting, as shown in the figure. In this case, seven different $\Lambda$ systems (three-level systems exhibiting EIT) contribute to the quantum memory, each with a slightly different energy difference. The energies of the created spin-waves $\hbar\omega_{sw} = \hbar\omega_s - \hbar\omega_g$ are shown in (c). 
}
\label{fig1}
\end{figure}

\begin{figure}
\includegraphics[width = \linewidth]{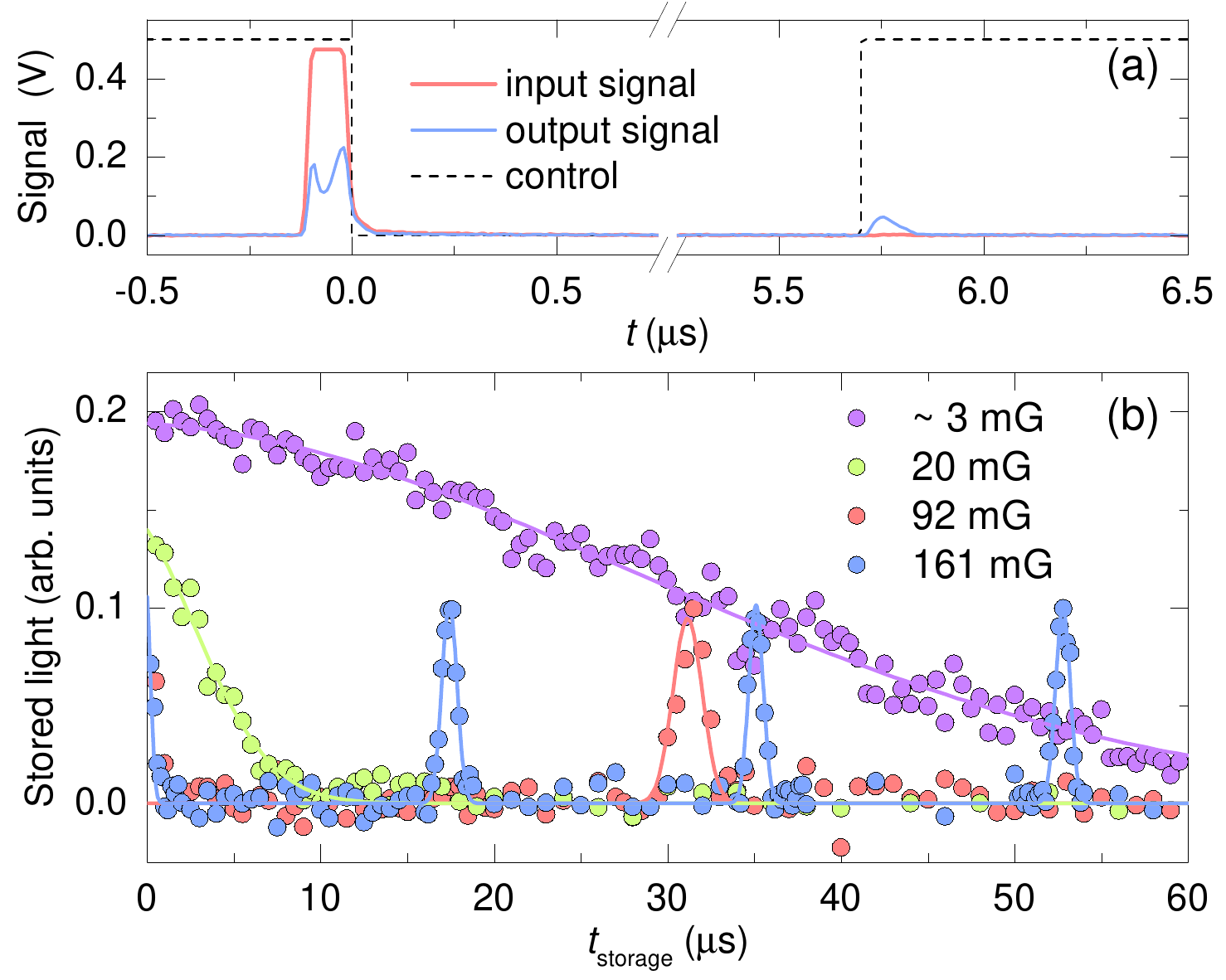}
\caption{
(a) A typical example of the measured signal from the photodiode. In red, we see the input signal pulse at the end of which the control beam also turns off. In blue, we see the leaked and stored pulse. The amount of stored light is determined by integrating the stored pulse.
(b) Retrieval efficiency with non-polarized atoms in magnetic fields of different strengths. Spin waves from different $m_F$ states interfere and create a periodic pattern in the light retrieval from the quantum memory as a function of storage time. The frequency of the occurrence of the revivals is proportional to the magnetic field strength. Around $t_{\mathrm{storage}} =$~\SI{53}{\micro s} we achieve a higher retrieval efficiency for \SI{161}{mG} than for the lowest achievable magnetic field. The lines are a guide to the eye.}
\label{fig2}
\end{figure}

\section{Experiment}

The experimental setup is illustrated in Fig. \ref{fig1}(a). We load the magneto-optical trap for \SI{7}{s} to prepare a cloud of $5 \times 10^7$ cesium atoms at $\sim$\SI{70}{\micro K}. Then we use the compressed MOT and molasses technique to further cool the atoms to $\sim$\SI{13}{\micro K} and transfer them into $F=3$. The details of this procedure can be found in Ref. \cite{meznarsic2019cesium}. We start the memory measurement \SI{4}{ms} after we turn off the MOT to ensure the quadrupole coil has completely turned off. The optical depth of the cloud is $\sim$10. Right before the memory measurement, we can shine a strong polarizing beam with $\sigma^+$ polarization on the $\vert F = 3\rangle \rightarrow \vert F' = 2\rangle$ transition. This transfers approximately 80\% of the atoms into $\vert F=3, m_F = 3 \rangle$. The quantization axis is defined by the magnetic field $B$ parallel to the signal beam.  

To store light, we shine on the atoms with the control beam and a \SI{0.5}{\micro s} pulse of the signal beam. The control beam drives the transition $\vert F = 4\rangle \rightarrow \vert F' = 4\rangle$ and the signal beam is on the $\vert F = 3 \rangle \rightarrow \vert F' = 4 \rangle$ transition as illustrated in Fig. \ref{fig1}(b). The power of the signal beam is a few microwatts and the power of the control beam is in the miliwatt range.Simultaneously with the end of the signal pulse, we turn off the control beam. After a selected storage time, we turn on the control beam again. We detect the stored signal light with a fast photodiode (Thorlabs PDA8A2, 50 MHz). The signal and the control light originate from two different diode lasers (Toptica DL pro and TA pro) that are offset phase locked using a beat note detector (Vescent D2-160 Beat Note Detector and D2-135 Offset Phase Lock Servo).

We shine the beams at a small angle of $\sim0.5^\circ$ that allows us to spatially filter the beams by blocking the control beam on an iris. In Section \ref{unpolarized}, we describe experiments where the signal beam is $\sigma^+$ polarized and the control is $\sigma^-$ polarized. In this case, the beams are additionally separated by a polarizing beam splitter. In experiments described in Section \ref{polarized}, both control and signal beam have $\sigma^+$ polarization, therefore we cannot use polarization filtering of the two beams.

We measure the efficiency of the quantum memory retrieval as a function of several parameters. A typical example of our measurement is shown in Fig. \ref{fig2}(a). On the photodiode measuring the transmittance of the signal beam, we observe two pulses - one at the time of the initial input, that is the leaked light, and another at the time when we turn the control beam back on, that is the stored light. We characterize the amount of stored light by integrating the stored light pulse.

\subsection{Unpolarized atoms} \label{unpolarized}

First, we demonstrate the occurrence of quantum memory revivals in a system of completely unpolarized atoms. In this experiment, we do not use the polarizing beam and the magnetic field is perpendicular to the direction of the probe beam. The control beam is, in this case, $\sigma^-$ polarized. All of this ensures that the atoms are distributed across all $m_F$ states. 

The amplitude of retrieved light pulses can be expressed as
\begin{equation}
    A(t) = A(0) \left \vert \sum_{n  = -3}^3 \sum_{m = -4}^4 \! P_{n, m} \mathrm{e}^{i\left(\omega_0 + (n + m) \frac{g\mu_B B}{\hbar}\right) t} \right \vert^2 \! f(t, \tau)\label{eq1}
\end{equation}
where $A(0)$ is the initial amplitude of the stored light (adapted from Ref. \cite{peters2009optimizing}) and $f(t, \tau)$ is a function describing the decay of efficiency due to dephasing with a characteristic lifetime $\tau$. The expression in the bracket of the exponential is the frequency of the spin-wave $\omega_{sw} = \omega_0 + (n+m)\frac{g\mu_B B}{\hbar}$, where we take into account that the ground state and storage state energies are split due to the Zeeman effect. The sums go over all 7 and 9 magnetic sublevels of the ground $F=3$ state and the $F=4$ states respectively. A combination of $n$ and $m$ corresponds to different coherences and their amplitudes are described by $P_{n, m}$. Because the coherences are formed by a two photon transition, $P_{n, m}$ is zero if $\vert n-m\vert > 2$, since the absorption or emission of a photon changes the magnetic number by at most one. $\omega_0$ is the frequency of the $F=3$ to $F=4$ clock transition in cesium, which is \SI{9.193}{GHz}. $\omega_L$ is the Larmor frequency $\omega_L = g\mu_B B/\hbar$, where $g \approx \SI{0.35}{MHz/G}$ is the Land\'{e} $g$-factor of $F=3$ ground state. In equation (\ref{eq1}) it is already taken into account that the Land\'{e} $g$-factor is of equal magnitude and opposite sign for $F=4$ ground state \cite{steck2019cesium}. $\mu_B$ is the Bohr magneton and $B$ is the applied magnetic field. The specific form of $f(t, \tau)$ depends on the dephasing mechanisms involved. Our results are best described by a Gaussian function $f(t, \tau) = \exp{(-t^2/\tau^2)}$.

We measure efficiency of the quantum memory retrieval as a function of the storage time for several different amplitudes of the magnetic field. The results are shown in Fig. \ref{fig2}(b). We observe that peaks in the retrieval efficiency occur every Larmor period $2\pi/\omega_L$. Interestingly, for longer storage times, the retrieval efficiency is higher at the peak of the memory revival at the highest shown magnetic field (\SI{161}{mG}, blue) than what we measure at the lowest achievable magnetic field (violet). From that, we conclude that even at the smallest magnetic field we can achieve, the magnetic field is not completely compensated and the observed lifetime $\tau=$\SI{44}{\micro s} is limited by magnetic dephasing rather than other mechanisms. 
The measurement at the higher magnetic field proves that the intrinsic lifetime of the memory is longer than the lifetime measured at the lowest achievable magnetic field. Since the widths of these peaks are inversely proportional to the magnitude of the magnetic field \cite{matsukevich2006observation}, we were able to approximate that there is $\sim\SI{3}{mG}$ of stray magnetic field present for the measurement shown in violet.

\subsection{Polarized atoms}\label{polarized}

We try to suppress the effect of the DSP collapses by polarizing the atomic cloud. Here we use the polarizing beam and set the polarization of the control and signal beam to both be $\sigma^+$, with the magnetic field in the direction of the beams. This way we excite fewer distinct spin waves, that are non-degenerate due to the Zeeman splitting. 

\begin{figure}
\centering
\includegraphics[width=\linewidth]{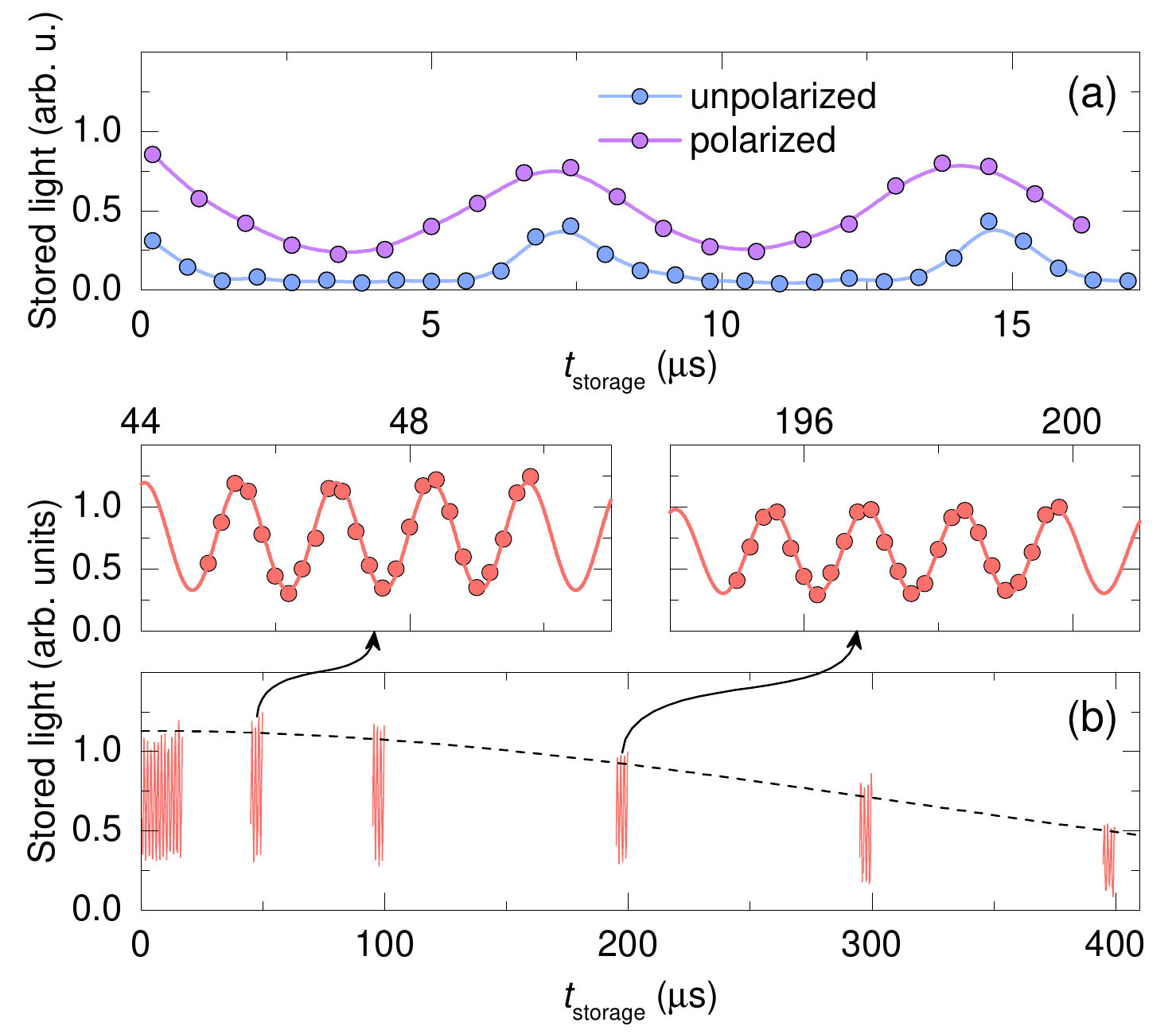}
\caption{The effect of polarizing the atoms and the lifetime of revivals. 
(a) Stored light as a function of storage time in a similar magnetic field with and without using a polarizing beam to polarize atoms.
(b) With  polarized atoms in an applied magnetic field we measured a lifetime of the quantum memory $\tau=$\SI{440}{\micro s} and without retrieval efficiency falling to zero for any storage time.
}
\label{fig3}
\end{figure}

To describe this situation, we rewrite Eq. (\ref{eq1}) by taking into account that we use $\sigma^+$-polarized signal and control beams. In this case, the only allowed coherences are ones with $m = n$ and their energies are equidistant with a difference of $2\omega_L$ as shown in Fig \ref{fig1}(c). This results in
\begin{equation}
    A(t) = A(0) \left \vert \sum_{m  = -3}^3 p_{m} \mathrm{e}^{i(\omega_0 + 2m \omega_L) t} \right \vert^2 f(t, \tau) \label{eq2}
\end{equation}
where we denoted $P_{m, m} = p_m$.  We see that, in this case, the revivals occur every half Larmor period and not only every Larmor period as in the previous section. 

Fig. \ref{fig3}(a) compares measurements in a similar magnetic field with and without the polarizing beam. We see that by using the polarizing beam we achieve that the retrieval efficiency no longer falls to zero between revivals. 

Fig. \ref{fig3}(b) shows a measurement of the intrinsic lifetime of the memory using polarized atoms in a magnetic field of \SI{1.0}{G}. The lifetime is $\tau = \SI{440}{\micro s}$. Insets show the oscillations of the retrieval efficiency for two different ranges of storage time.

\begin{figure}
\includegraphics[width = \linewidth]{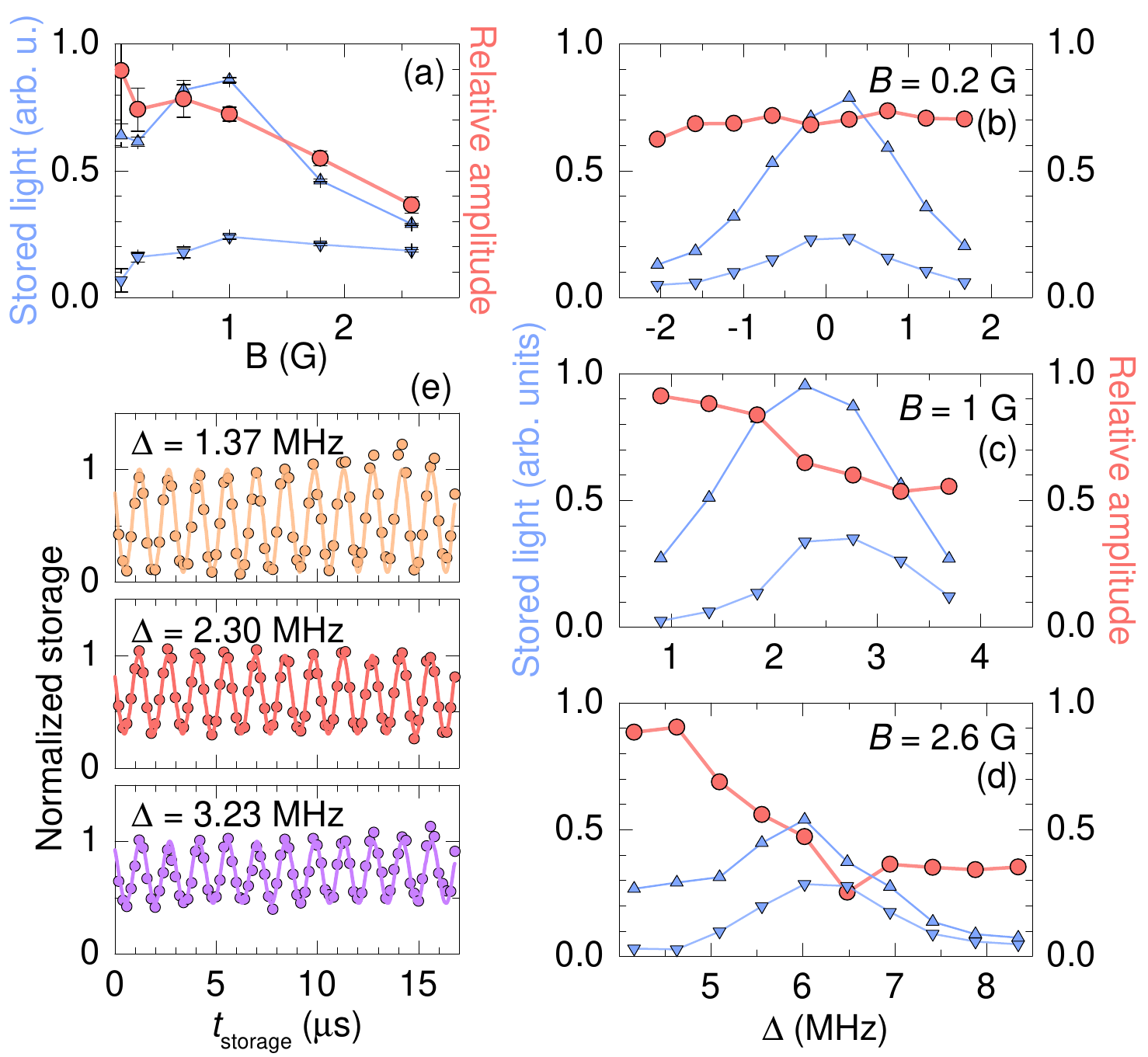}
\caption{
Dependence of the variation of the retrieval efficiency of the quantum memory as a function of the applied magnetic field and the detuning of the control beam. Upwards pointing blue triangles show the maximal retrieval efficiency of storage and downwards pointing blue triangles show the minimal retrieval efficiency. The red circles show the relative amplitude of the oscillations. 
(a) The dependence on the magnetic field where the data is measured at such detuning that peak retrieval efficiency is maximal. The ratio clearly shows suppression of the DSP collapses for larger magnetic fields.
(b-d) show the dependence of retrieval efficiency of the quantum memory on the detuning $\Delta$ for three  different magnetic fields. 
As the frequency difference is increased, the process becomes selective for the transitions that are most changed by the magnetic field (highest $m_F$). This decreases the amplitude of DSP collapses. 
(e) Shows the oscillations of the retrieval efficiency for the first \SI{17}{\micro s} of storage time for three different detunings. We show normalized retrieval efficiency, which is the retrieval efficiency divided by the maximum. 
}
\label{fig4}
\end{figure}

We measure the amplitude of the oscillations as a function of the strength of the magnetic field. In Fig. \ref{fig4}(a) we plot the maximal and the minimal retrieval efficiency and the relative amplitude of oscillations. The oscillations are clearly suppressed at higher magnetic fields.

We see that in the experiment, time evolution of the retrieved light has a simple one-frequency cosine function shape. This is because only two coherences are formed -- $p_3$ and $p_2$, as the atoms are almost perfectly polarized to $m_F = 3$, with a small percentage of atoms still left in $m_F = 2$. In this case, equation (\ref{eq2}) simplifies to
\begin{equation}
     A(t) = A(0) \left[ p_3^2 + p_2^2 + 2p_2 p_3 \cos(2\omega_Lt) \right] f(t, \tau)
\end{equation}
The maximal retrieved signal (without decay) is therefore $A_{max} = A(0) (p_3+p_2)^2$ and the peak-to-peak amplitude of the oscillations is ${A_{osc}=A(0)4p_2p_3}$. The ratio of these two, denoted by $R$, represents the relative amplitude of the oscillations, and it is equal to 
 \begin{equation}
     R = A_{osc}/A_{max} = 4p_2(1-p_2), \label{eq4}
 \end{equation} where we already took into account that $p_2 + p_3 = 1$. 

In measurements shown to this point, the frequencies of the signal and control beam were set to the value that resulted in the highest peak efficiencies. Here, we present another way we can make the memory more selective for one $m_F$ component, that is, by increasing the frequency difference of the control and signal beam. In zero magnetic field, the frequency difference for this transition is $\omega_0$. We describe the frequency detuning with the $\Delta = \omega_{\mathrm{sig}} - \omega_{\mathrm{con}} - \omega_0$, where $\omega_{\mathrm{sig}}$ and $\omega_{\mathrm{con}}$ are the frequencies of the signal beam and the control beam respectively. Fig. \ref{fig1}(c) shows how the frequency of the spin wave $\omega_{sw}$ depends on $m_F$ and the magnetic field, and we expect that the frequency difference of the beams should follow $\omega_{sw}$.

We measure the highest and the lowest point of the oscillations as a function of $\Delta$. We select the frequency of the signal beam that results in the highest retrieval efficiency and scan the $\Delta$ by only changing the frequency of the control beam. The results are shown in Fig. \ref{fig4}(b-d) for \SI{0.2}{G}, \SI{1.0}{G} and \SI{2.6}{G}. In the first case, where $B=\SI{0.2}{G}$, the relative amplitude of the oscillations is independent on the frequency and the peak of the maximal retrieval efficiency is centered on $\Delta = 0$. However, for \SI{1}{G}, the peak in the maximal retrieval efficiency is at $\Delta = $ \SI{2.2}{MHz} and the relative amplitude significantly decreases for higher $\Delta$. Examples of the oscillations of retrieval efficiency at \SI{1}{G} for three different detunings are shown in Fig. \ref{fig4}(e). We see a similar effect for \SI{2.6}{G}, leading to even lower relative amplitudes that plateau for higher detunings. The position of the peak of the maximal retrieval efficiency aligns well with the shift due to the Zeeman splitting $\Delta_{sw} = \omega_{sw} - \omega_0 = 2 g_F \mu_B B m_F$ of $m_F = 3$, which is \SI{2.1}{MHz} for \SI{1.0}{G} and \SI{5.5}{MHz} for \SI{2.6}{G}. The smallest relative amplitude we observed is $R = 0.25$, measured at $\Delta = \SI{6.5}{MHz}$ at \SI{2.6}{G}. Using Equation (\ref{eq4}) we can calculate that this corresponds to $p_2 = 0.07$, meaning that 93\% of light is stored in the coherence with $m_F=3$. 

\section{Discussion and Conclusions}

It is very challenging to completely annul the magnetic field in a cold atom experiment where the use of magnetic shields is not possible. This means that the effective lifetime of a quantum memory is often limited by the residual magnetic field and not the intrinsic lifetime of the memory. We show that it is beneficial to instead add a strong magnetic field and polarize the atoms into predominantly one $m_F$ state. 
This way we were able to show the intrinsic lifetime in our system is $\tau=\SI{440}{\micro s}$, even though it is otherwise limited to \SI{44}{\micro s} by the stray magnetic field. From the shape of the decay of efficiency $f(t,\tau)$, which is Gaussian, we conclude that the dominant decoherence mechanism in our system are magnetic gradients. $\tau=\SI{440}{\micro s}$ corresponds to a gradient of $\sim\SI{7}{mG/cm}$ \cite{veissier2013quantum}, which agrees with our previous assessment of magnetic field gradients in our system \cite{gosar2021single}. From the temperature of the atoms and the angle between the beams, we estimate that if we could eliminate the effect of these gradients, the lifetime would be limited to $\sim$\SI{700}{\micro s} by atomic motion.

The main requirement for this experiment is that the magnetic field is in the direction of the beams and that both the signal and the control beam are $\sigma^+$ polarized. This way, the coherence is formed from states $\vert g \rangle$ and $\vert s \rangle$ with the same $m_F$ (as shown in Fig. \ref{fig1}(b)). The number of different coherences is therefore much lower than when the magnetic field is perpendicular to the direction of the beams and the coherences form between states with $m_F$ and $m_F \pm 1$ as well. We decrease the number of populated coherences by polarizing the atoms towards the stretched $m_F$ state with a pulse of the polarizing beam before performing the quantum memory experiment. Additionally, we show that we can decrease the influence of $m_F = 2$ by detuning the control beam  towards higher frequencies, as it then becomes selective for the highest $m_F$ state.   With optical pumping techniques, the initial MOT could be prepared in a single $m_F$ state, in which case the collapses and revivals can be avoided completely \cite{zhao2009long, xin2019transporting, dudin2013light}.

In this paper, we show stored light and the retrieval efficiency in arbitrary units, because the input power of the control and signal beams varies between experimental runs. However, it should be noted that in our system with optical depth of $\sim$10, the efficiencies of light pulse storage reach up to 7\%. This could be improved by using an elongated and denser MOT, leading to higher optical depth and therefore higher efficiency \cite{vernaz2018highly, hsiao2014cold}. 

Even though the collapses and revivals of dark-state polaritons present a challenge when trying to achieve a continuous quantum memory, their presence could be useful for certain types of storage. For example, we could exploit the revivals for time-multiplexing of the quantum memory \cite{farrera2018entanglement}. In principle, one could send two signal pulses into the same atomic cloud  and read the signals at the time of the corresponding revival for each input pulse separately. Here, the complete collapse of the dark-state polariton would ensure that the output pulse would consist of purely the corresponding input, since the other input is completely suppressed.

Additionally, one can imagine that the presence of different possible coherences would allow for multiplexing in the frequency of the signal and control beam. The writing process requires the difference between the frequency of the control and the signal beam to correspond  to the atomic transition and the width of this process is only in the MHz range, much narrower than the individual atomic transition. In a high enough magnetic field, the EIT resonances for each Zeeman sublevel would be separated by more than the width of the EIT and we could excite the coherences separately. 

\begin{acknowledgments}

We thank Rok \v{Z}itko, Jure Pirman, Ticijana Ban and Damir Aumiler for their comments and discussions.
This work was supported by the Slovenian Research Agency (Research Core Fundings No. P1-0125, No. P1-0099 and No. P1-0416, and Research Project No. J2-2514).

K.G. and V.P.J. contributed equally to this work.

\end{acknowledgments}

\bibliography{revivals}

\end{document}